\documentclass[12pt]{article}
\usepackage{graphicx}
\usepackage{epsfig}
\begin{document}
\title{Open Flavour Charmed Mesons in a QCD Potential model}
\author{ $^{1}$Krishna Kingkar Pathak and $^{2}$D K Choudhury \\
$^{1}$Deptt. of Physics,Arya Vidyapeeth College,Guwahati-781016,India\\
e-mail:kkingkar@gmail.com\\
$^{2}$Deptt. of Physics, Gauhati University, Guwahati-781014,India\\
and\\
Physics Academy of Northeast, Guwahati-781014}
\date{}
\maketitle
\begin{abstract}
 
We modify the mesonic wavefunction by using a short distance scale $ r_{0}$ in analogy with Hydrogen atom and estimate the values of masses and decay constants of the open flavour charm mesons $D$, $D_{s}$ and $B_{c}$ within the framework of a QCD Potential Model. We also calculate leptonic decay widths of these mesons to study branching ratios and life time. The results are in good agreement with experimental and other theoretical values.   
\end{abstract}
\section{Introduction}
 Recently, we have reported a regularisation procedure to avoid the singularity by introducing a flavour dependent short distance scale  at the origin to study the oscillation frequency of $B$ and $\overline{B}$ mesons in a QCD Potential Model[1]. The purpose of this letter is to use the Potential Model to calculate the masses and decay constants of open flavour charmed mesons $D$, $D_{s}$ and $B_{c}$ and then to find the decay width and branching ratio of the same.\\
If the CKM element is well known from other
measurements, then $ f_{p}$ can be well measured. If, on the other
hand, the CKM element is less well or poorly measured, having
theoretical input on $ f_{P}$  can allow a determination of the CKM
element.
These decay constants can be accessed both experimentally and through lattice Quantum Chromodynamics (lQCD) simulations. While for $f_{\pi}$ , $f_{K}$ , $f_{D}$ , experimental measurements agree well
with lattice QCD calculations, a discrepancy is seen for the value of $f_{D_{s}}$ : The 2008 PDG average
for $f_{D_{s}}$ is $273 \pm 10 MeV $[2], about $3\sigma$ larger than the most precise $N_{f}= 2 + 1$ lQCD result from
the HPQCD/UKQCD collaboration [3], $241 \pm 3$ MeV. On the other hand, experiments and lQCD
calculations agree very well with each other on the value of $f_{D}$ , $f_{D} (expt)= 205.8 \pm 8.9$ MeV and
$f_{D} (lQCD)= 207 \pm 4$ MeV. The discrepancy concerning $f_{D_{s}}$ is quite puzzling because whatever
systematic errors have affected the lQCD calculation of $f_{D}$ , they should also be expected for the
calculation of $f_{D_{s}}$ [4]. \\
However, the discrepancy is reduced to $2.4\sigma$ with the new (updated) data from CLEO [5,6] and Babar [7], together with the Belle measurement [8] and the latest PDG average is $f_{D_{s}}=257.5 \pm 6.1$ MeV [9]. Lately the HPQCD collaboration has
also updated its study of the $D_{s}$ decay constant [10]. By including additional results at smaller
lattice spacing along with improved determinations of the lattice spacing and improved tuning of
the charm and strange quark masses, a new value for the $D_{s}$ decay constant has been reported as $f_{D_{s}} = 248.0 \pm 2.5$ MeV.
  A better understanding about the decay properties of $D$ and $D_{s}$ in this regard is quite important.\\  
The properties of the $B_{c}$ meson are of special interest \cite{11}, since it is the only heavy meson consisting of two heavy quarks with different flavours. A pecularity of $B_{c}$ decays is that both the quarks may involve in its weak decays.From experimental point of view, study of weak decays of $B_{c}$ meson is quite important for the determination of CKM elements. More detailed information about its decay properties are expected in near future at LHC and other experiments. In this letter the decay constants of $D$, $D_{s}$ and $B_{c}$ are computed and are used to study the decay width and decay time of the same. \\
Rest of the paper is organised as follows:
  In section 2, we discuss the formalism, while in section 3 we summarise the calculations and results. Section 4 contains summary and conclusion.\\
	    
\section{Formalism}
\subsection{Wave function in the model}

The wave function in a specific potential Model with its relativistic corection procedure is \cite{1,12}:
\begin{equation}
\psi_{rel+conf}\left(r^{\prime}\right)=\frac{N^{\prime}}{\sqrt{\pi a_{0}^{3}}} e^{\frac{-r^{\prime}}{a_{0}}}\left( C^{\prime}-\frac{\mu b a_{0}(r^{\prime})^{2}}{2}\right)\left(\frac{r^{\prime}}{a_{0}}\right)^{-\epsilon}
\end{equation}
where
\begin{equation}
N^{\prime}=\frac{2^{\frac{1}{2}}}{\sqrt{\left(2^{2\epsilon} \Gamma\left(3-2\epsilon\right) C^{\prime 2}-\frac{1}{4}\mu b a_{0}^{3}\Gamma\left(5-2\epsilon\right)C^{\prime}+\frac{1}{64}\mu^{2} b^{2} a_{0}^{6}\Gamma\left(7-2\epsilon\right)\right)}}
\end{equation}
\begin{equation}
C^{\prime}=1+cA_{0}\sqrt{\pi a_{0}^{3}}
\end{equation}
\begin{equation}
\mu=\frac{m_{i}m_{j}}{m_{i}+m_{j}}
\end{equation}
\begin{equation}
a_{0}=\left(\frac{4}{3}\mu \alpha_{s}\right)^{-1}
\end{equation}
\begin{equation}
\epsilon=1-\sqrt{1-\left(\frac{4}{3}\alpha_{s}\right)^{2}}
\end{equation}
\begin{equation}
r^{\prime}=r+r_{0}
\end{equation}
and the QCD potential is taken as
\begin{equation}
V\left(r\right)=-\frac{4}{3r}\alpha_{s}+br+c
\end{equation}

\subsection{The mesonic wave function at the origin}
The term $\left(\frac{r+r_{0}}{a_{0}}\right)^{-\epsilon}$ in eq.1 is used to incorporate relativistic effect in a free Dirac way [12,16]. The cut off parameter is used to regularise the wavefunction at the origin such that the wavefunction remain approximately constant from $r_{0}$ down to $r=0$. In analogy to QED, we use the cut off parameter to regularise the wavefunction as[1]\\
\begin{equation}
r_{0}\sim a_{0}e^{-\frac{1}{\epsilon}}.
\end{equation}
Unlike QED, it is flavour dependent depending on the flavours of the quark masses. 

 The input parameters in the numerical calculations are used as $m_{u}=0.336 GeV$, $m_{s}=0.483 GeV$, $m_{c}=1.550 GeV$ and $m_{b}=4.950 GeV$, $b=0.183GeV^{2}$,$c=-0.5 GeV$, $\alpha_{s}=0.39$ and  $\alpha_{s}=0.22$ at charm and bottom mass scale.The value of the undetermined factor $A_{0}$, which appears in the series solution[12] is so chosen that the consistancy of the value of $cA_{0}=1 GeV^{3/2}$ is sustained with our previus works[1,13]. With these values, we compute the short distance scale $r_{0}$ for open flavour charmed mesons $D$, $D_{s}$ and $B_{c}$ and is shown in table 1. \\
\label{tab1}
\begin{table}
\begin{center}
\caption{Values of cut-off parameter $r_{0}$ for different mesons }
\vspace {.2in}
\begin{tabular}{|c|c|c|}\hline
mesons&reduced mass & {values of $r_{0}$ in $ Gev^{-1}$}\\\hline 

$D(c\bar{u}/c\bar{d})$&0.2761&0.0073\\\hline
$D_{s}(c\bar{s})$&0.3682&0.0055\\\hline

$B_{c}(\bar{b}c)$&1.1838&$3.872\times 10^{-10}$\\\hline

\end{tabular}
\end{center}
\end{table}
It shows as the masses of the mesons increase, $r_{0}$ is decreased. For heavy-heavy mesons like $B_{c}$ it becomes as tiny as $10^{-10}GeV^{-1}$ which presumably infers that the relativistic effect is too small for this heavy meson.\\

\section{Calculation and results}
\subsection{ masses and decay constants of open flavoured charm mesons}
 The decay constant with relativistic correction can be expressed
through the meson wave function $\Phi_{P}(p)$ in the momentum
space \cite{17,18}as\\
\begin{equation}
f_{P} =\sqrt{\frac{12}{M_{P}}}\int\frac{d^3p}{(2\pi)^3} \left(
\frac{E_q + m_q}{2E_q}\right)^{1/2} \left(\frac{E_{\bar
q} + m_{\bar q}}{2E_{\bar q}}\right)^{1/2}\left(1+
\lambda_{P} \frac{p^2}{[E_q+m_q][E_{\bar q}+m_{\bar
q}]}\right)\Phi_{P}(p)
\end{equation}
 with $\lambda_P=-1$ . In the nonrelativistic limit $\frac{p^2}{m^2}<<1.0$, this expression reduces to the well
known relation between  $f_{P}$ and the ground state wave function at the origin $\psi\left(0\right)$ the Van-Royen-Weisskopf formula \cite{19}.

\begin{equation}
f_{p}=\sqrt{ \frac{12}{M_{p}}|\psi\left(0\right)|^{2}}
\end{equation}
with QCD correction factor the decay constant can be written as[20]
\begin{equation}
f_{p}=\sqrt{ \frac{12}{M_{p}}|\psi\left(0\right)|^{2}\overline{C}^{2}},  where  \overline{C}^{2}=1-\alpha_{s}/\pi[2-\frac{m_{Q}-m_{\overline Q}}{m_{Q}+m_{\overline Q}}ln\frac{m_{Q}}{m_{\overline Q}}]
\end{equation}
where $M_{p}$ is the pseudoscalar meson mass in the ground state and can be obtained as[28]
\begin{equation} 
M_{p}=m_{Q}+m_{\overline Q}+
\langle H \rangle
\end{equation}
where
\begin{equation} 
\langle H \rangle= \langle \frac{p^{2}}{2\mu}\rangle +\langle V\left(r\right)\rangle
\end{equation}

In table.2 and 3 we record the prediction of the model for masses and decay constants.\\
\begin{table}
\begin{center}
\caption{The masses of heavy-light  mesons in GeV.}
\begin{tabular}{|c| c| c| }\hline

mesons & masses & experimental values\\\hline

$D(c\bar{u}/c\bar{d})$&1.860&$1.869\pm0.0016$\cite{9} \\\hline
$D(c\bar{s})$&1.959&$1.968\pm 0.0033$\cite{9}\\\hline
$B_{c}(\bar{b}c)$&6.507&$6.277\pm0.006$\cite{9}\\\hline

\end{tabular}
\end{center}
\end{table}
\begin{table}
\begin{center}
\caption{The decay constants of heavy-light  mesons
 with and without QCD correction in GeV}
\begin{tabular}{|c| c| c| c| c| }\hline
&  \multicolumn{2}{c|}{our work }& other work \\
meson  & \multicolumn{2}{c|}{}  &   \\\hline
 & \multicolumn{1}{l}{$f_{p}$} &$ f_{pc}$ & \\

& \multicolumn{1}{l}{ }&  &\\\hline
$D(c\bar{u}/c\bar{d})$&\multicolumn{1}{l|}{0.240}&0.225 &$0.205\pm0.085\pm 0.025$ \cite{21}[Exp]\\
&  &  & $0.206 $ \cite{22} \\
&  &  & 0.234 [23]\\
&  &  & 0.208 [24]\\
&  &  & 0.201 [25]\\
&  &  & 0.235 [26]\\\hline
$D(c\bar{s})$&\multicolumn{1}{l|}{0.291}&0.266 & $0.254\pm 0.059$ \cite{21}[Exp]\\
&  &  & $0.245$ \cite{22}\\
&  &  & 0.268 [23]\\
&  &  & 0.256 [24]\\
&  &  & 0.249 [25]\\
&  &  & 0.266 [26]\\\hline

$B_{c}(\bar{b}c)$&\multicolumn{1}{l|}{0.435}&0.413 & 0.433 \cite{27}[Theory]\\
 &  & &0.470[28]& \\\hline
\end{tabular}\\
$f_{p}$= decay constants without QCD correction\\
                        
$f_{pc}$= decay constant with QCD correction\\ 
\end{center}
\end{table}
\subsection{ Leptonic decay width and Branching ratio of D and $D_{s}$ mesons}
Charged mesons formed from a quark and anti-quark can decay to a charged lepton pair when these objects annihilate via a virtual $W^{\pm}$ boson. Quark-antiquark annihilations via a virtual $W^{+}(W^{-})$ to the $l^{+}\nu(l^{-}\nu$ final states occur for the $\pi^{\pm}, K^{\pm}, D^{\pm}_{s} and B^{\pm}$ mesons. There are several reasons for studying the purely leptonic decays of charged mesons. Such processes are rare but they have clear experimental signatures due to the presence of a highly energetic lepton in the final state. The theoretical predictions are very clean due to the absence of hadrons in the final state. The total leptonic decay width of D, $D_{s}$ mesons are given by

\begin{equation}
\Gamma \left(D^{+}_{q}\rightarrow{l^{+}\nu}\right)= \frac{ G^{2}_{F}|V_{cq}|^{2}f^{2}_{D_{q}}}{8\pi}m^{2}_{l}\left(1-\frac{m^{2}_{l}}{M^{2}_{D_{q}}}\right)^{2} M_{D_{q}},   q=d,s
\end{equation}

 These transitions are helicity suppressed; i.e., the amplitude is proportional to $m_{l}$, the mass of the lepton l, in complete analogy to  $\pi\rightarrow {l^{+}\nu}$.\\
The leptonic widths of the charged D and $D_{s}$ mesons are obtained using equation.12,  employing the predicted values of the pseudoscalar decay constants $f_{D}$ and $f_{s}$ along with the masses of the $M_{D}$ and $M_{D_{s}}$ from our work. The leptonic widths for seperate lepton channel by the choice of $m_{l=\tau,\mu,e}$ are computed. Branching ratio of $D_{q}$ mesons are calculated by using the relation
\begin{equation}
BR=\Gamma \times \tau
\end{equation}
The life time of these mesons are $\tau_{D}=1.04 ps$ and $\tau_{D_{s}}=0.5 ps$ are taken from the world average value reported by Particle data group (PDG-2010)[9]. The present results as tabulated in table.4 are in accordance with the available experimental values.
\begin{table}
\begin{center}
\caption{The leptonic branching ratio of D and $D_{s}$ mesons.}
\begin{tabular}{|c|c|c|c|c|}\hline
mesons&$ BR_{\tau}\times 10^{-3}$ &$ BR_{\mu}\times 10^{-4}$ & $ BR_{e}\times 10^{-8}$\\\hline
 
$D(c\bar{u}/c\bar{d})$&0.78(0.68) &4.4(3.89) &1.05(0.92)  \\

Expt.[9]&$< 1.2$ &$3.82$ & \\
B. Patel etal.,[29]&0.9&6.6&1.5\\\hline
&$ BR_{\tau} \times 10^{-2}$ &$ BR_{\mu}\times 10^{-3}$ & $ BR_{e}\times 10^{-7}$\\\hline
$D(c\bar{s})$&6.4(5.3) &7.0(5.9) &1.67(1.40) \\

Expt.[9]&$ 5.6\pm0.4$ &$5.8\pm0.4$ & \\
B.Patel etal.,[29]&8.4&7.7&1.8\\\hline

\end{tabular}
 
\end{center}
*Values within the bracket represent the branching ratio for $f_{p}$ with QCD correction.
\end{table}
\subsection{ Weak decay of $B^{+}_{c}$ meson}
Adopting the spectator model for the charm beauty mesons system [30], the total decay width of $B^{+}_{c}$ meson can be approximated as the sum of the widths of $\bar{b}$-quark decay keeping c-quark as spectator, the c-quark decay with $\bar{b}$ as spectator, and the annihilation channel $B^{+}_{c}\rightarrow {l^{+}\nu_{l}(c\bar{s},u\bar{s}),l=e,\mu,\tau}$ with no interference assumed between them.
Acordingly, the total width is written as [30]
\begin{equation}
\Gamma \left(B_{c}\rightarrow {X}\right)=\Gamma \left(b\rightarrow {X}\right) +\Gamma \left(c\rightarrow {X}\right)+\Gamma \left(anni\right)
\end{equation}
Neglecting the quark binding effects, the b and c inclusive widths in the spectator approximation are[30]
\begin{equation}
\Gamma \left(b\rightarrow {X}\right)= \frac{9 G^{2}_{F}|V_{cb}|^{2}m^{5}_{b}}{192\pi^{3}}
\end{equation}
\begin{equation}
\Gamma \left(c\rightarrow {X}\right)= \frac{5 G^{2}_{F}|V_{cs}|^{2}m^{5}_{c}}{192\pi^{3}}
\end{equation}
Here we have used the model quark masses and the CKM matrix elements $|V_{cs}|=0.957$, $|V_{cb}|=0.039$ from the particle data group.\\
Employing the computed mass and pseudoscaler decay constant from the present study, the width of the annihilation channel is computed using the expression given by [28,30],
\begin{equation}
\Gamma \left(Anni\right)= \frac{ G^{2}_{F}|V_{bc}|^{2}f^{2}_{B_{c}}M_{B_{C}}}{8\pi}m^{2}_{q}\left(1-\frac{m^{2}_{q}}{M^{2}_{B_{c}}}\right)^{2} C_{q}
\end{equation}
Where $C_{q}=1$ for the$\tau \nu_{\tau}$ channel and $C_{q}=3|V_{cs}|^2$ for $c\bar{s}$ channel, and $m_{q}$ is the mass of the heaviest fermions.The computed results of the annihilation decay rate and total decay rate are tabulated in table.5. Our prediction for life time with these results lie well within the experimental value and is placed in table.5.
\begin{table}
\begin{center}
\caption{Decay width(in $10^{-4}eV$) and life time of $B_{c}$ meson}
\begin{tabular}{|c|c|c|c|}\hline
meson&$ \Gamma \left(Anni\right)$ &$ \Gamma \left(B_{c}\rightarrow {X}\right)$ & $ \tau  \left( inps\right) $\\\hline

$B_{c\bar{b}}$&1.17(1.06) &19.17(19.06) &0.344(0.346)  \\\hline
[9]& & & $0.453\pm0.041$\\\hline
[30] &1.40& 14.00&0.47\\\hline
[31]&0.67&8.8&0.75\\\hline
\end{tabular}
\end{center}
\end{table}

\section{Conclusion}
In the present work, we have computed the masses of heavy-light mesons in a specific potential model and use a short distance scale to regularise the wave function near the origin to compute the decay constants. The short distance scale as well as the decay parameters are very sensitive to the strong coupling constant $\alpha_{s}$. It is to be noted that strong coupling constant as is used in ref.[1] cannot be used for $D$, $D_{s}$ and $B_{c}$ mesons since the results oversoots the experimental and theoretical values and hence the model is excluded to treat the $D$ and $B$ mesons together. However we use the same $\alpha_{s}$ as is used in our previous work[13,14,15] and with a different renormalisation mass scale one can regain these strong coupling constants within the formula of ref.[1]. \\ 

The analysis with QCD correction  is found to be closer to experiments and other theoretical results. In our calculation we have found  $\frac{f_{D_{s}}}{f_{D}}$=1.18 (with QCD correction) and $\frac{f_{D_{s}}}{f_{D}}$=1.21 (without QCD correction), which is in accordance with the lattest QCD Sum rule result $1.193\pm 0.025\pm 0.007$ \cite{22}, PDG average $\frac{f_{D_{s}}}{f_{D}}=1.25\pm 0.06\pm$[9], as well as with the recent lattice results
 $f_{D_{s}} /{f_{D}} = 1.164 \pm 0.011$ [3,10] and $f_{D_{s}} /{f_{D}} = 1.20 \pm 0.02$ [32].\\
 Present study on the leptonic decay branching ratio of D and $D_{s}$ mesons with QCD correction for $\tau$ and $\mu$ leptonic channels presented in Table.4 are as per the available experimental limits. Large experimental uncertainity in the electron channel make it difficult for any reasonable conclusion. The computed result within the framework of the QCD potential model for annihilation decay width as well as the lifetime of $B_{c}$ meson is also found to be  well with the available datas as presnted in the Table.5. Probaly, in future high luminosity better statistics and high confidence level data sets will be able to provide more light on the spectroscopy and decay properties of these open charm mesons.  \\

\section*{Acknowledgement}
One of the author, K K Pathak is grateful to Suvankar Roy from Gauhati University for fruitful discussions and comments.

\section{Appendix}
Equations of calculation for kinetic energy, coulomb and confining terms of the potential to obtain masses of the pseudoscalar mesons.

\begin{eqnarray}
\langle H \rangle=\langle \frac{p^{2}}{2\mu}\rangle +\langle V(r)\rangle \nonumber& \\
\\
\langle \frac{p^{2}}{2\mu}\rangle= -\frac{1}{2\mu}   (   \frac{2\pi C_{1}^{2}\epsilon \left(\epsilon-1\right)}{\left(2g\right)^{1-2\epsilon}}\Gamma\left(1-2\epsilon\right)-\frac{2\pi h_{1} C_{1}\epsilon \left(\epsilon-1\right)}{\left(2g\right)^{3-2\epsilon}}\Gamma\left(3-2\epsilon\right)
+\frac{4\pi gC_{1}^{2} \left(\epsilon+1\right)}{\left(2g\right)^{2-2\epsilon}}\Gamma\left(2-2\epsilon\right)\nonumber\\
-\frac{4\pi gh_{1}C_{1} \left(\epsilon+1\right)}{\left(2g\right)^{4-2\epsilon}}\Gamma\left(4-2\epsilon\right)
-\frac{2\pi C_{1} \left(g^{2}C_{1}+h_{1}(2-\epsilon)(3-\epsilon) \right)}{\left(2g\right)^{3-2\epsilon}}\Gamma\left(3-2\epsilon\right)+\nonumber\\
\frac{2\pi h_{1} \left(g^{2}C_{1}+h_{1}(2-\epsilon)(3-\epsilon) \right)}{\left(2g\right)^{5-2\epsilon}}\Gamma\left(5-2\epsilon\right)+\frac{4\pi gh_{1}C_{1} }{\left(2g\right)^{4-2\epsilon}}\Gamma\left(4-2\epsilon\right)\nonumber\\-\frac{4\pi gh^{2}_{1}}{\left(2g\right)^{6-2\epsilon}}\Gamma\left(6-2\epsilon\right)+\frac{2\pi C_{1} g^{2}h_{1}}{\left(2g\right)^{5-2\epsilon}}\Gamma\left(5-2\epsilon\right)-\frac{2\pi g^{2}h^{2}_{1}}{\left(2g\right)^{7-2\epsilon}}\Gamma\left(7-2\epsilon\right)
)  
\nonumber\\
\\
\langle \frac{-\alpha_{c}}{r}\rangle=-\alpha_{c}2\pi C_{1}^{2}\frac{1}{(2g)^{2-2\epsilon}}\Gamma\left(2-2\epsilon\right)-\alpha_{c}2\pi h_{1}^{2}\frac{1}{(2g)^{6-2\epsilon}}\Gamma\left(6-2\epsilon\right)+\alpha_{c}4\pi h_{1} C_{1}\frac{1}{(2g)^{4-2\epsilon}}\Gamma\left(4-2\epsilon\right)\nonumber\\
\\
\langle{br}\rangle=b2\pi C_{1}^{2}\frac{1}{(2g)^{4-2\epsilon}}\Gamma\left(4-2\epsilon\right)+b2\pi h_{1}^{2}\frac{1}{(2g)^{8-2\epsilon}}\Gamma\left(8-2\epsilon\right)-b4\pi h_{1} C_{1}\frac{1}{(2g)^{6-2\epsilon}}\Gamma\left(6-2\epsilon\right)\nonumber\\
\\
where
h_{1}=hk_{1}\nonumber\\
h=\frac{\mu b a_{0}}{2}\nonumber\\
k_{1}=\frac{k}{l}\nonumber\\
l=\frac{1}{a_{0}^{-\epsilon}}\nonumber\\
k=\frac{N^{\prime}}{(\pi a_{0})^{\frac{1}{2}}}\nonumber\\
C_{1}=C^{\prime}k_{1}\nonumber\\
g=\frac{1}{a_{0}}\nonumber\\
\alpha_{c}=4\alpha_{s}/3\nonumber\\
\nonumber\\&
\end{eqnarray}

\begin{thebibliography}{99}
\bibitem{1} Pathak K K and Choudhury D K; Chinese Physics Letter. Vol. 28,No.10,101201(2011) \\
\bibitem{2} Amsler C et al. (Particle data group), Phys. Lett.B,1(2008)\\
\bibitem{3} Follana E, Davies C.T.H, Lepage G. P. and  Shigemitsu J. [HPQCD Collaboration and UKQCD
Collaboration], Phys. Rev. Lett. 100, 062002 (2008).\\
\bibitem{4} Geng L. S., Altenbuchinger M. and Weise W.; arXiv:hep-ph/1012.0666(2010)\\
\bibitem{5} Onyisi P.U.E et al. [CLEO Collaboration], Phys. Rev. D 79, 052002 (2009).\\
\bibitem{6} Naik P. et al. [CLEO Collaboration], Phys. Rev. D 80, 112004 (2009).\\
\bibitem{7} Lee J. P. et al [The BABAR Collaboration], arXiv:1003.3063 [hep-ex].\\
\bibitem{8} Widhalm L. et al. [Belle Collaboration], Phys. Rev. Lett. 100, 241801 (2008).\\
\bibitem{9} Nakamura K. et al. [Particle Data Group], J. Phys. G 37, 075021 (2010)\\
\bibitem{10} Davies C.T.H, McNeile C., Follana E., Lepage G.P, Na H. and Shigemitsu J. Phys. Rev. D82, 114504 (2010)\\
\bibitem{11} CDF collaboration, Corcoran M.D,(2005);hep-ex/050601; Abulencia A. et al.,Phys.Rev.Lett.96(2006)082002\\
\bibitem{12} Choudhury D K, Das P,Goswami D D and Sarma J K ; Pramana J.Phys.46, 349(1996) \\
\bibitem{13} Hajarika B J, Pathak K K and Choudhury D K; MPLA. Vol. 26, No. 21 (2011) 1547-1554\\
\bibitem{14} Choudhury D K and Bordoloi N S ; MPLA,Vol.17, No.29; 1909(2002)\\
\bibitem{15} Choudhury D K and Bordoloi N S; MPLA, vol. 24; 443(2009)\\
\bibitem{16} Itzykson C  and Zuber J  in ``Quantum Field Theory'';(International Student Edition , McGraw Hill ,Singapore ,1986),pp-79\\
\bibitem{17} Galkin V.O., Mishurov A.Yu. and Faustov R.N.; Sov. J. Nucl. Phys.53; 1026(1991) \\
\bibitem{18} Vinodkumar P C, Rai A K and Patel B; arXiv: hep-ph/0808.1776v1, Frascati Physics series Vol. XLVI(2007)\\
\bibitem{19} Van Royen R. et al., Nuovo Cimento 50, (1967).\\
\bibitem{20} E. Braaten and S. Fleming Phys. Rev D 52 (1995) 181\\
\bibitem{21} Asner D. et al. (Heavy Flavor Averaging Group), eprint arXiv:1010.1589.\\
\bibitem{22} Wolfgang Lucha, Dmitri Melikhov,Silvano Simula; arxiV:hep-ph/1101.5986,arXiv:hep-ph/1108.0844(2011)\\
\bibitem{23} Ebert D et al.,Phys. Lett.B634, 214 (2006)\\
\bibitem{24} Mao-Zhi Yang; arXiv:hep-ph/1104.3819(2011)\\
\bibitem{25} Aubin C et. al, Phys. Rev. Lett.95,122002 (2005)\\
\bibitem{26} Chiu T W et al., Phys. Lett.B624,31 (2005)\\
\bibitem{27} Ebert D, Faustov R.N. and Galkin V.O.,Phys.Rev D 67 014027(2003)\\
\bibitem{28} Rai A K, Patel B and Vinodkumar P C, Phys. Rev. C78, 055202(2008)\\
\bibitem{29} Patel B. and Vinodkumar P.C., arXiv:hep-ph/0908.2212v1(2009)\\
\bibitem{30} El-Hady A.Adb,Lodhi M.A.K and Vary J.P,Phys.Rev.D 59 094001(1999)\\
\bibitem{31} Godfrey S.,Phys.Rev.D 70,054017(2004)\\
\bibitem{32} Bazavov A. et al.(Fermilab Lattice and MILC Collaborations),PoS LAT2009, 249 (2009)\\.

\end{thebibliography}
\end{document}